\documentclass[11pt]{article}
\pdfoutput=1
\usepackage{aas_macros,jcappub,natbib,float,caption,comment,bm}
\usepackage{graphicx,epsfig}
\usepackage[dvipsnames]{xcolor}
\usepackage[utf8]{inputenc}

\def\be{\begin{equation}}
\def\ee{\end{equation}}
\def\ba#1\ea{\begin{align}#1\end{align}}
\newcommand{\vs}{\nonumber\\}

\let\oldv\v

\renewcommand{\v}[1]{\mathbf{#1}}

\newcommand{\vr}{\v{r}}	
\newcommand{\vk}{\v{k}}

\newcommand{\refeq}[1]{eq.~(\ref{eq:#1})}
\newcommand{\refeqs}[2]{eqs.~(\ref{eq:#1})--(\ref{eq:#2})}
\newcommand{\reffig}[1]{figure~\ref{fig:#1}}
\newcommand{\refFig}[1]{Figure~\ref{fig:#1}}
\newcommand{\refsec}[1]{Section~\ref{sec:#1}}

\renewcommand{\[}{\left[}
\renewcommand{\]}{\right]}
\renewcommand{\(}{\left(}
\renewcommand{\)}{\right)}

\newcommand{\hMpc}{~h^{-1}~{\rm Mpc}}
\newcommand{\ihMpc}{~h~{\rm Mpc}^{-1}}
\newcommand{\lya}{\text{Lyman-}\alpha}
\renewcommand{\d}{\delta}
\newcommand{\para}{\parallel}
\newcommand{\dnl}{{\mathcal D}_{\rm NL}}

\newcommand{\Dz}{\Delta z}
\newcommand{\sinc}{{\rm sinc}}
\newcommand{\tr}{\tilde{r}}

\title{The Lyman-$\alpha$ power spectrum - CMB lensing convergence cross-correlation}
\author[a]{Chi-Ting Chiang,}
\author[b]{and An\oldv{z}e Slosar}
\affiliation[a]{C.N. Yang Institute for Theoretical Physics, Stony Brook University, Stony Brook, NY 11794, U.S.A.}
\affiliation[b]{Brookhaven National Laboratory, Blgd 510, Upton, NY 11375, U.S.A.}
\emailAdd{chi-ting.chiang@stonybrook.edu,anze@bnl.gov}

\abstract{We investigate the three-point correlation between the $\lya$
  forest and the CMB weak lensing ($\delta_F \delta_F \kappa$) expressed
  as the cross-correlation between the CMB weak lensing field
  and local variations in the forest power spectrum. In addition to
  the standard gravitational bispectrum term, we note the existence of
  a non-standard systematic term coming from mis-estimation of the mean flux
  over the finite length of $\lya$ skewers. We numerically calculate the
  angular cross-power spectrum and discuss its features. We integrate
  it into zero-lag correlation function and compare our predictions
  with recent results by Doux et al.. We find that our predictions are
  statistically consistent with the measurement, and including the systematic
  term improves the agreement with the measurement. We comment on the
  implication of the response of the $\lya$ forest power spectrum to the
  long-wavelength density perturbations.}

\begin{document}

\subheader{\rm YITP-SB-17-30}

\maketitle
\flushbottom

%%%%%%%%%%%%%%%%%%%%%%%%%%%%%%%%%%%%%%%%%%%%%%%%%%%
\section{Introduction}
\label{sec:introduction}
The three-point function of the large-scale structure contains ample information
that cannot be probed by the two-point function, hence it is sensitive to the
nonlinear gravitational evolution \cite{Bernardeau:2001qr}, tracer bias
\cite{Desjacques:2016}, and even the inflationary physics \cite{Alvarez:2014vva}.
Measurement of the three-point function, or its Fourier counterpart bispectrum,
by auto- or cross-correlation is thus one of the biggest scientific goals for ongoing
and future surveys.

In this paper, we propose a new bispectrum formed by the $\lya$ forest power spectrum
(hereafter forest power spectrum) and the CMB lensing convergence. Since the forest power
spectrum is sensitive to the small-scale matter fluctuation at $2\le z\le4$ whereas
CMB lensing convergence measures the total matter fluctuation along the line-of-sight,
this combination of bispectrum allows us to probe the nonlinear structure formation
that has not been explored using galaxies as tracers. Note that a similar observable
has been measured in Ref.~\cite{Doux:2016xhg} by cross-correlating the one-dimensional
forest power spectrum and the CMB lensing at {\it the same} angular position.
We shall generalize the bispectrum to include the complete angular scale between
the $\lya$ forest skewers and the CMB lensing, and demonstrate that their observable
is an integral over the angular scale of the total bispectrum.

The rest of this paper is organized as follows.
In \refsec{true_signal}, we compute the underlying signal generated by the gravitational evolution.
In \refsec{biased_signal}, we calculate the signal due to the continuum fitting when
measuring the $\lya$ forest, which we refer to as the ``continuum-misestimation bias''.
In \refsec{observation}, we discuss the contamination from damped $\lya$ absorbers
and compare our prediction with the measurement in Ref.~\cite{Doux:2016xhg}.
We conclude in \refsec{conclusion}.
Throughout the paper we adopt the Planck cosmology \cite{Ade:2015xua},
i.e. $h=0.6803$, $\Omega_bh^2=0.0226$, $\Omega_ch^2=0.1186$, $A_s=2.137\times10^{-9}$,
and $n_s=0.9667$.

%%%%%%%%%%%%%%%%%%%%%%%%%%%%%%%%%%%%%%%%%%%%%%%%%%%
\section{Underlying signal due to gravitational evolution}
\label{sec:true_signal}
The mean of the forest power spectrum at redshift $z$ with width $\Dz$,
corresponding to the comoving distance $r_\para$ and width $\Delta r$,
can be written as \cite{McDonald:2001fe}
\be
 P_{FF}(\vk_s)=P_{FF}(k_s,\mu_s)
 =b_F^2(1+\beta_F\mu_s^2)^2P_{\d\d}(k_s)\dnl(k_s,\mu_s) \,,
\label{eq:pff_fid}
\ee
where $\vk_s$ is the small-scale wavevector (opposed to the large-scale mode
that we introduce later), $\mu_s$ is the cosine of $\vk_s$ along the line-of-sight,
$b_F$ is the flux bias, $\beta_F$ is the redshift-space distortion parameter
for the flux, $P_{\d\d}$ is the linear power spectrum, and $\dnl$ is the fitting
formula for the small-scale nonlinearity. In this paper we shall use the fitting
function provided in Ref.~\cite{Arinyo-i-Prats:2015vqa} as
\be
 \dnl(k_s,\mu_s)=\exp\Bigg\lbrace\[q_1\Delta^2(k_s)+q_2\Delta^4(k_s)\]
 \[1-\(\frac{k_s}{k_v}\)^{a_v}\mu_s^{b_v}\]-\(\frac{k_s}{k_p}\)^2\Bigg\rbrace \,,
\label{eq:dnl}
\ee
where $\Delta^2(k)=k^3P_{\d\d}(k)/(2\pi^2)$ and $(q_1,q_2,k_v,a_v,b_v,k_p)$
are fitting parameters obtained from simulations. Note that the linear power
spectrum as well as all biases and fitting parameters (provided in Table 4
and 5 of Ref.~\cite{Arinyo-i-Prats:2015vqa}) depend on redshift, but we do
not write the dependence explicitly when no confusion occurs.

If there is a large-scale fluctuation on the sky, the forest power spectrum would be modulated
and at the leading order the forest power spectrum at angular position $\vr_\perp$
becomes \cite{Chiang:2017vsq}
\ba
 P_{FF}(\vk_s,\vr_\perp)=P_{FF}(\vk_s)+\frac{dP_{FF}(\vk_s)}{d\bar\d}\bar\d(\vr_\perp) \,,
\ea
where $\bar\d$ is the two-dimensional projected large-scale density fluctuation,
and $dP_{FF}(\vk_s)/d\bar\d$ is the response of the forest power spectrum with
respect to $\bar\d$ due to the gravitational evolution, which can be measured from
the separate universe simulations \cite{Cieplak:2015kra,Chiang:2017vsq}. In the
flat-sky approximation, the three-dimensional position can be decomposed into
parallel $r_\para$ and transverse $\vr_\perp$ components, and $\bar\d$ is given by
\be
 \bar\d(\vr_\perp)
 =\frac{1}{\Delta r}\int_{r_\para-\Delta r/2}^{r_\para+\Delta r/2}
 dr'_\para\delta(r'_\para,\vr_\perp)
 =\frac{1}{\Delta r}\int dr'_\para\delta(r'_\para,\vr_\perp)
 \Theta\(\frac{\Delta r}{2}-|r'_\para-r_\para|\) \,,
\label{eq:bard}
\ee where $\delta$ is the underlying matter density perturbation and
$\Theta$ is the heaviside step function. At the leading order, i.e in
the limit where the wavevector of the large-scale mode $\bar\d$ is much
smaller than $\vk_s$, the response depends only on the small-scale mode
$\vk_s$ and is independent of the wavelength of $\bar\d$.

Consider measuring the correlation between the forest power spectrum with
the two-dimensional projected CMB lensing convergence in the same redshift
bin
\ba
 \kappa(\vr_\perp)\:&=\int_{r_\para-\Delta r/2}^{r_\para+\Delta r/2}dr'_\para
 \int d^2r'_\perp\delta(\vr')W_\kappa(r'_\para)\Lambda_\kappa(\vr'_\perp-\vr_\perp) \vs
 \:&=\int d^3r'\delta(\vr')W_\kappa(r'_\para)\Theta\(\frac{\Delta r}{2}-|r'_\para-r_\para|\)
 \Lambda_\kappa(\vr'_\perp-\vr_\perp) \,,
\label{eq:kappa}
\ea
where $W_\kappa$ is the lensing kernel and $\Lambda_\kappa$ is the
weighting filter in the transverse direction (which we will later set
to be the Wiener filter). The lensing kernel is given by
\be
 W_\kappa(r_\para)=\frac{3H_0^2\Omega_m}{2c^2}\frac{r_\para}{a(r_\para)}\frac{r_*-r_\para}{r_*} \,,
\ee
where $H_0$ is the present-day Hubble, $\Omega_m$ is the fractional energy
density of matter, $c$ is the speed of light, $a$ is the scale factor,
and $r_*$ is the comoving distance to the last-scattering surface. The
three-point correlation between $P_{FF}$ and $\kappa$ in Fourier space
is thus
\be
 \langle P_{FF}(\vk_s,\vk_{l\perp})\kappa(\vk'_{l\perp})\rangle'
 =\frac{dP_{FF}(\vk_s)}{d\bar\d}\langle\bar\d(\vk_{l\perp})\kappa(\vk'_{l\perp})\rangle'
 =\frac{dP_{FF}(\vk_s)}{d\bar\d}P^{\rm 2d}_{\bar\d\kappa}(\vk_{l\perp}) \,,
\label{eq:p2d_pffkappa}
\ee
where $\langle\rangle'$ is the ensemble average without the Dirac delta function.
Using \refeqs{bard}{kappa}, we can compute $P^{\rm 2d}_{\bar\d\kappa}$ as
\ba
 P^{\rm 2d}_{\bar\d\kappa}(\vk_{l\perp})\:&=
 \int_{r_\para-\Delta r/2}^{r_\para+\Delta r/2}dr'_{1\para}W_\kappa(r'_{1\para})
 \int dr'_{2\para}\frac{1}{\Delta r}\Theta\(\frac{\Delta r}{2}-|r'_{2\para}-r_\para|\) \vs
 \:&\hspace{1cm}\times\int\frac{dq_\para}{2\pi}P_{\d\d}(q_\para,\vk_{l\perp})
 \Lambda_\kappa(-\vk_{l\perp})e^{iq_\para(r'_{1\para}-r'_{2\para})} \vs
 \:&=\int_{r_\para-\Delta r/2}^{r_\para+\Delta r/2}dr'_\para W_\kappa(r'_\para)
 \int\frac{dq_\para}{2\pi}P_{\d\d}(q_\para,\vk_{l\perp})\Lambda_\kappa(-\vk_{l\perp})
 \sinc\(\frac{q_\para\Delta r}{2}\)e^{iq_\para(r'_\para-r_\para)} \,,
\label{eq:p2d_bardkappa}
\ea
and {\it the only assumption we make is that $P_{\d\d}$ is independent of
$z$ within $\Delta z$}. Because of the sinc function, the contribution to
$P^{\rm 2d}_{\bar\d\kappa}$ from the parallel mode is mainly from $q_\para\lesssim\Delta r^{-1}$.
Physically, this means that $P^{\rm 2d}_{\bar\d\kappa}$ is sensitive to
the largest parallel mode, which is set by the redshift bin size. Following
Ref.~\cite{LoVerde:2008re}, in the flat-sky approximation we can convert the
two-dimensional power spectrum to the angular power spectrum as
\be
 l(l+1)C_l^{\bar\d\kappa}\approx k^2_{l\perp}P^{\rm 2d}_{\bar\d\kappa}(\vk_{l\perp}) \,,
\label{eq:cl_bardkappa}
\ee
where $l+1/2\approx r_\para k_{l\perp}$. Combining \refeqs{p2d_pffkappa}{cl_bardkappa},
we have the most general correlation between forest power spectrum and CMB
lensing convergence, which includes both the dependence on the small-scale
mode $\vk_s$ and the large-scale mode $\vk_{l\perp}$, or equivalently the
angular scale $l$. This calculation is general and can be used for other
fields in place of Lyman-$\alpha$, most notably 21-cm
emission from intensity mapping.

\begin{figure}[t]
\centering
\includegraphics[width=0.495\textwidth]{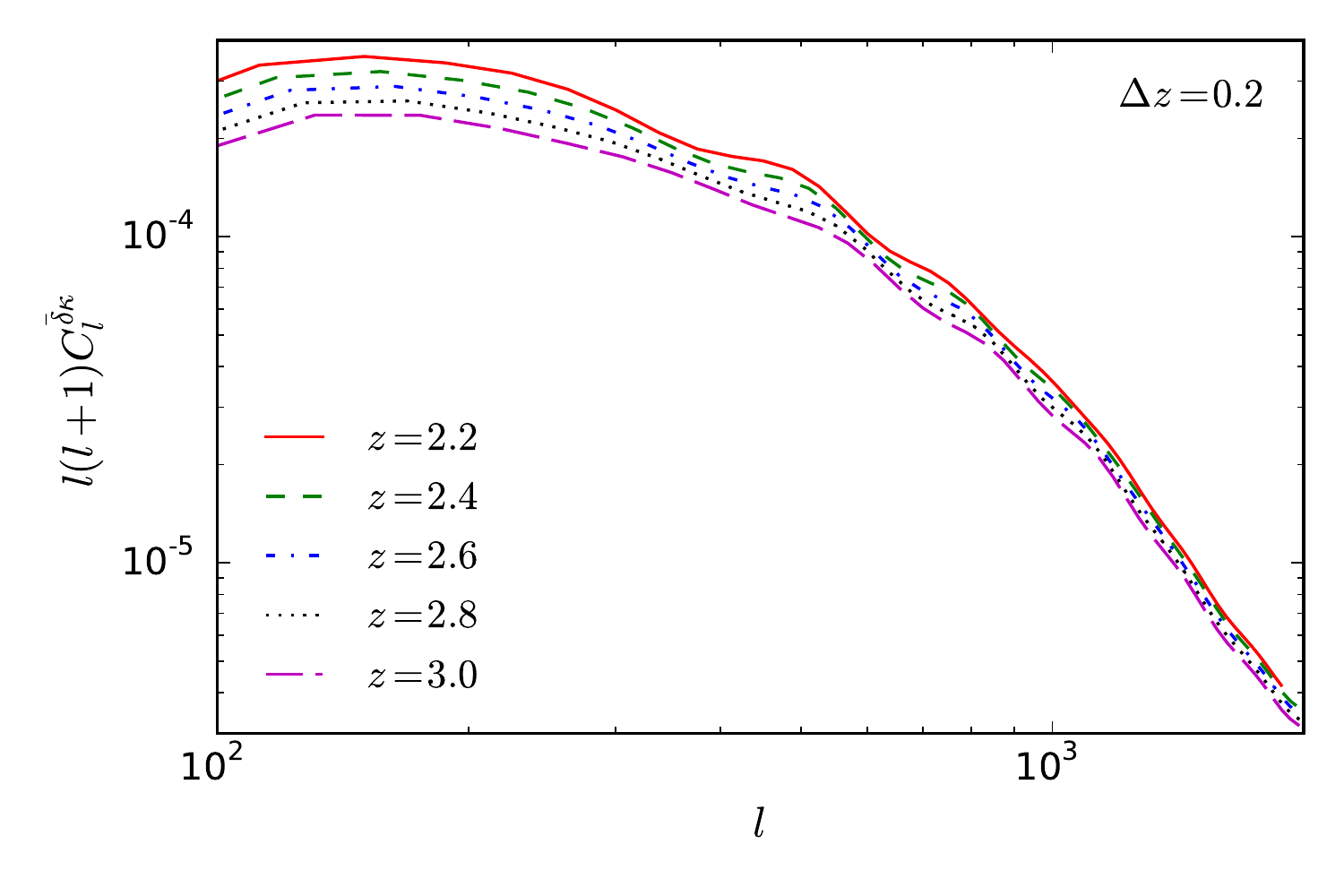}
\includegraphics[width=0.495\textwidth]{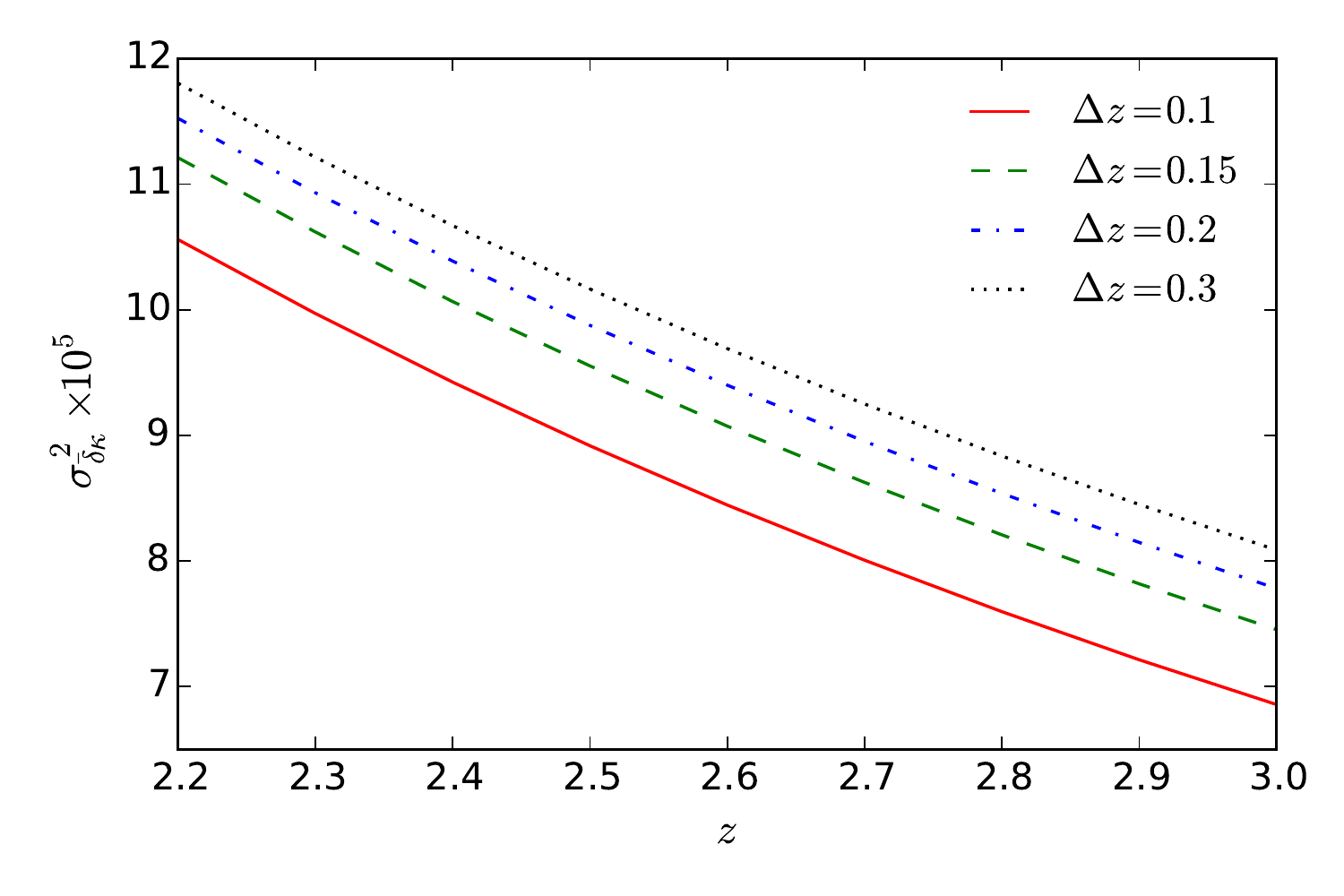}
\caption{(Left) The angular power spectrum $C_l^{\bar\d\kappa}$ with $\Delta z=0.2$
as a function of the angular scale $l$ at various redshifts denoted by different
colors and line styles. We set $l\ge100$ so that the flat-sky approximation works
well. (Right) The correlation of $\bar\d$ and $\kappa$ at the same angular position,
$\sigma^2_{\bar\d\kappa}$, as a function of redshift for various $\Delta z$ denoted
by different colors and line styles.}
\label{fig:corr_bardkappa}
\end{figure}

The left panel of \reffig{corr_bardkappa} shows $C_l^{\bar\d\kappa}$ with
$\Delta z=0.2$ as a function of the angular scale $l$ at various redshifts.
We set $l\ge100$ so that the flat-sky approximation works well. To numerically
evaluate $C_l^{\bar\d\kappa}$, we set $q_{\para,\rm max}=1000\ihMpc$ and
confirm that the result is insensitive to the choice of $q_{\para,\rm max}$.
We also take $W_\kappa$ to be the Wiener filter from Ref.~\cite{Ade:2015zua} with $l=r'_\para k_{l\perp}$.
The power spectrum is larger at lower redshift, and this is the result of
the gravitational evolution. To check the approximation that the linear power
spectrum does not vary within the redshift bin $\Delta z$, we include the
redshift evolution and find that the fractional difference is less than 1\%.
The agreement is slightly better at higher redshift since $\Delta r$ is smaller
at larger $z$ when $\Delta z$ is fixed.

\begin{figure}[t]
\centering
\includegraphics[width=0.495\textwidth]{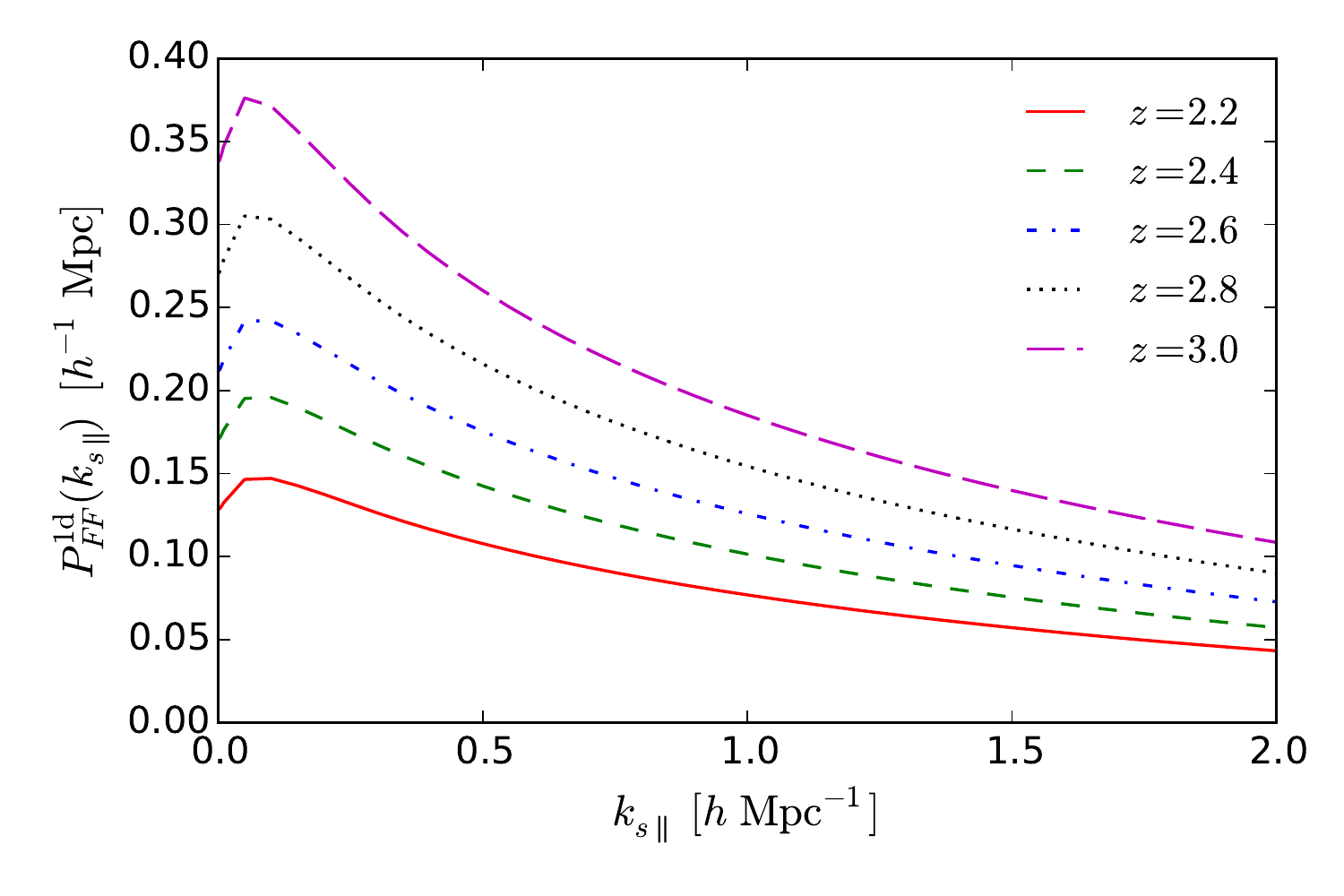}
\includegraphics[width=0.495\textwidth]{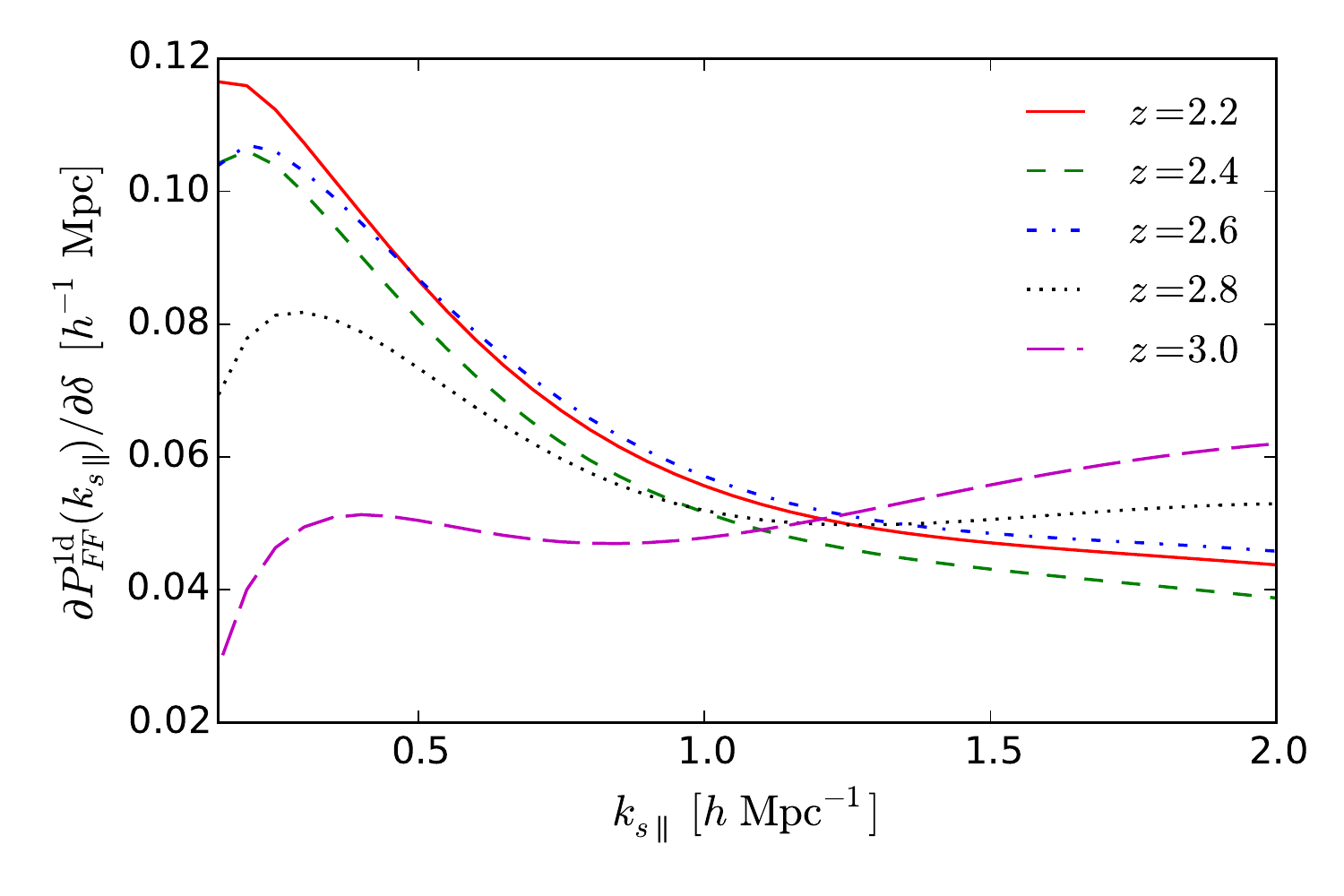}
\caption{(Left) One-dimensional $\lya$ forest power spectrum at different redshifts.
(Right) Response of the one-dimensional $\lya$ forest power spectrum to long-wavelength
overdensity evaluated with the separate universe simulations and \refeq{pff1d_resp}.}
\label{fig:pff1d}
\end{figure}

Since the forest power spectrum is frequently measured as the one-dimensional
power spectrum, we can Fourier transform in the small-scale transverse direction:
\be
 P^{\rm X}_{FF}(k_{s\para},\vr_{s\perp})=\int\frac{d^2k_{s\perp}}{(2\pi)^2}P_{FF}(k_{s\para},\vk_{s\perp})
 e^{i\vk_{s\perp}\cdot\vr_{s\perp}},
\ee
with the one-dimensional power spectrum given by
$P_{FF}^{\rm 1d}(k_{s\para})=P^{\rm X}_{FF}\left(k_{s,\para},\vr_{s\perp}=0\right)$.
Similarly one can Fourier transform the cross-correlation of \refeq{p2d_pffkappa}
in the small-scale transverse wavevector $\vk_{s\perp}$ to get a cross-correlation
as a function of ($k_{s\para},\vr_{s\perp},k_{l\perp})$. This tells us how a
cross-power spectrum between two close skewers as a function of the small-scale
parallel wavevector $k_{s\para}$ and small-scale perpendicular separation $\vr_{s\perp}$
responds to a large-scale mode in $\kappa$ field fluctuation. For this we need
\ba
 \frac{dP^{\rm X}_{FF}(k_{s\para},\vr_{s\perp})}{d\bar\d}
 \:&=\int\frac{d^2k_{s\perp}}{(2\pi)^2}\frac{dP_{FF}(k_{s\para},\vk_{s\perp})}{d\bar\d}
 e^{i\vk_{s\perp}\cdot\vr_{s\perp}} \vs
 \:&=\int\frac{d^2k_{s\perp}}{(2\pi)^2}P_{FF}(k_{s\para},\vk_{s\perp})
 \frac{d\ln P_{FF}(k_{s\para},\vk_{s\perp})}{d\bar\d} e^{i\vk_{s\perp}\cdot\vr_{s\perp}} \,.
\label{eq:pff1d_resp}
\ea
As before, the one-dimensional specialization is just the case when we set
$\vr_{s\perp}=0$, i.e.
$dP^{\rm 1d}_{FF}(k_{s\para})/d\bar\d=dP^{\rm X}_{FF}(k_{s\para},\vr_{s\perp}=0)/d\bar\d$.

To numerically evaluate the one-dimensional forest power spectrum, we use
\refeqs{pff_fid}{dnl} and set $k_{s\perp,\rm max}=1000\ihMpc$. To compute
the response of the one-dimensional forest power spectrum, we first measure
and smooth $d\ln P_{FF}(k_{s\para},\vk_{s\perp})/d\bar{\d}$ from the separate 
universe simulations as described in Refs.~\cite{Cieplak:2015kra,Chiang:2017vsq}
and then use the analytic form of $P_{FF}(k_{s\para},\vk_{s\perp})$, i.e.
\refeqs{pff_fid}{dnl}. To minimize
the possible bias on the integration from the smoothing and interpolation, we
set $k_{s\perp,\rm max}=10\ihMpc$. \refFig{pff1d} shows the one-dimensional forest
power spectrum (left) and its response to $\bar\d$ (right). The one-dimensional
forest power spectrum is larger at higher redshift due to the increase of $|b_F|$.
Conversely, the response of the one-dimensional forest power spectrum, which
contains the responses from linear power spectrum, flux bias, and the nonlinear
fitting function, is larger at lower redshift for $k\lesssim1\ihMpc$.

\begin{figure}[t]
\centering
\includegraphics[width=0.495\textwidth]{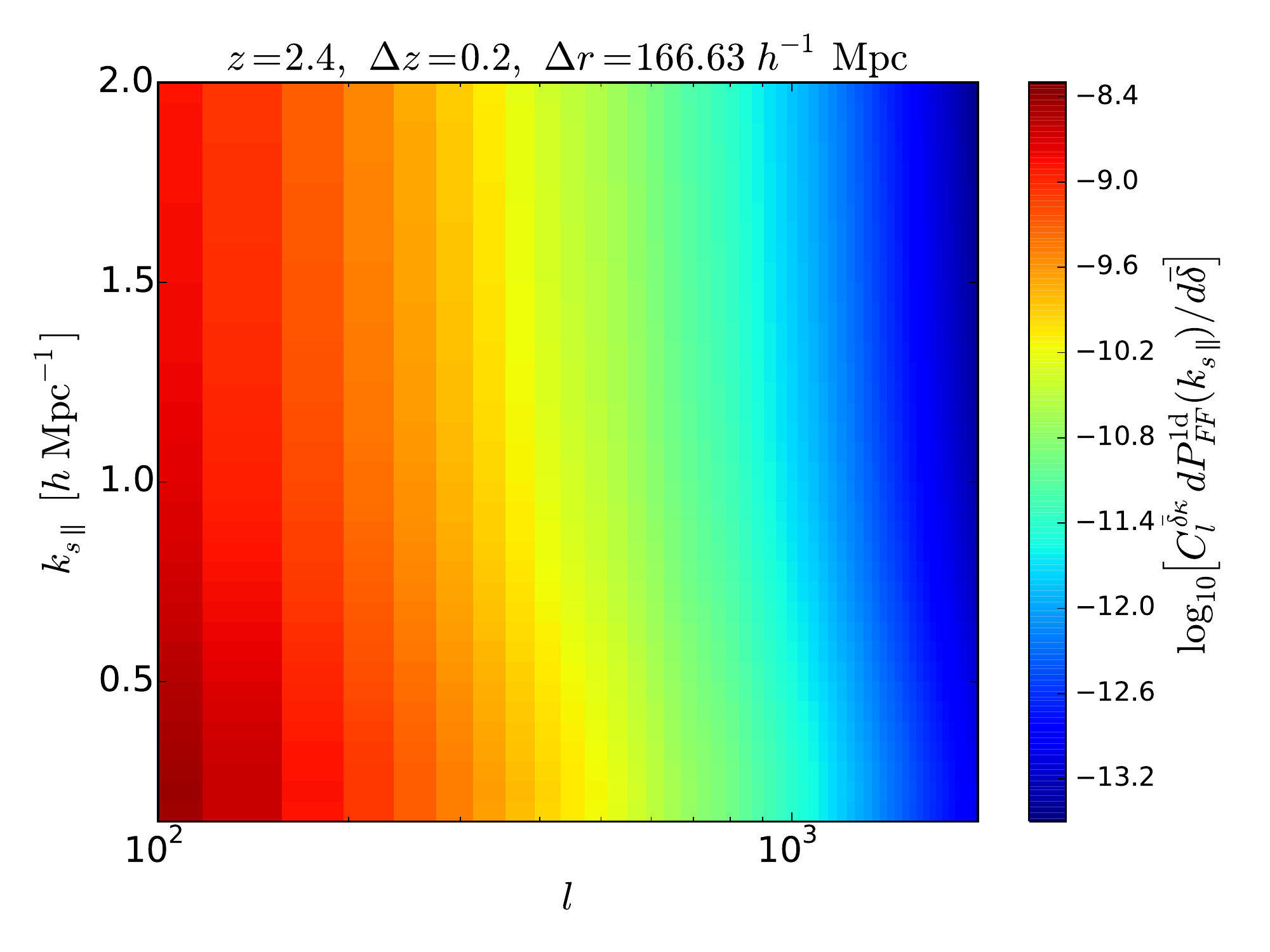}
\includegraphics[width=0.495\textwidth]{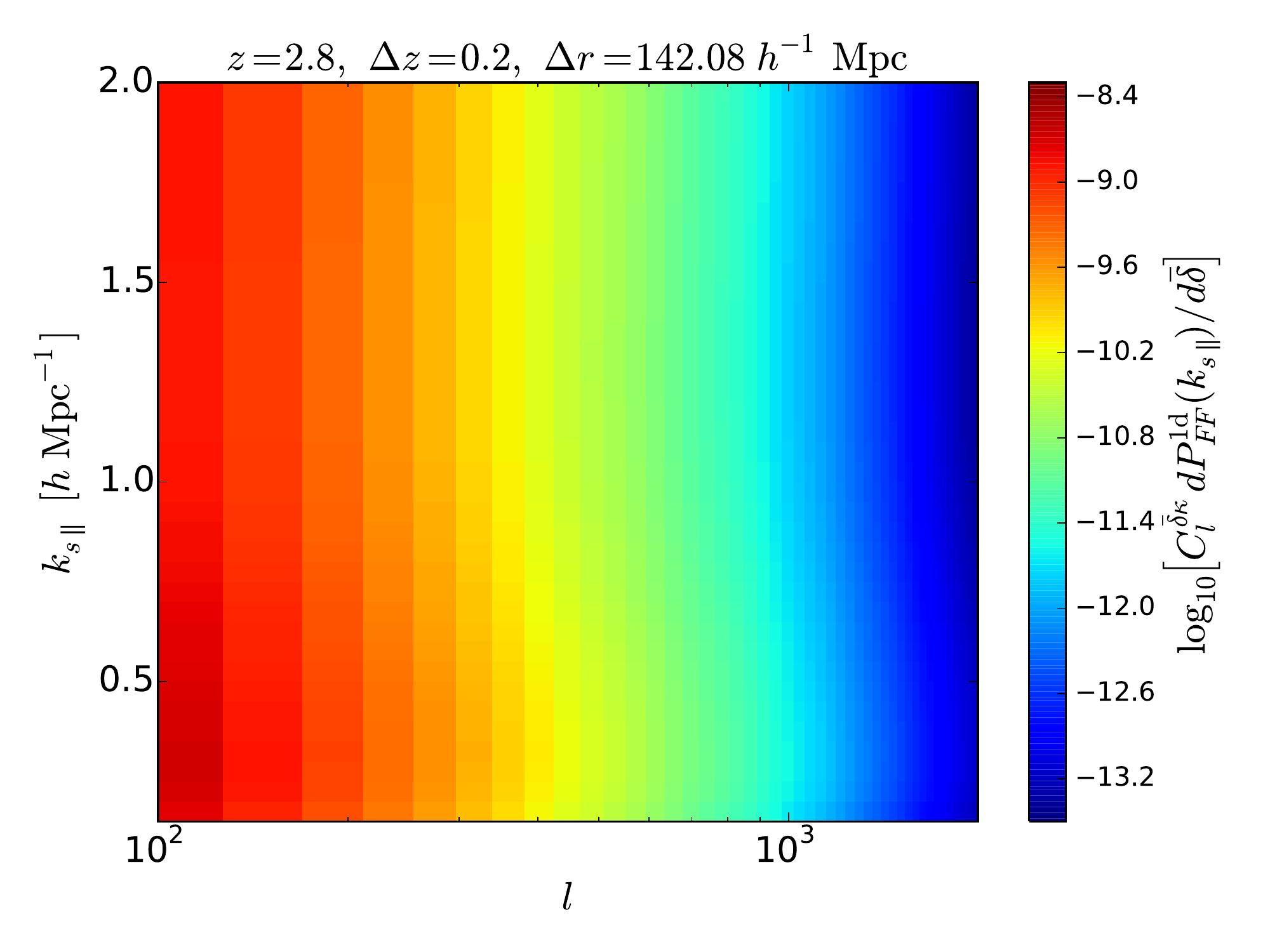}
\caption{Correlation of the one-dimensional $\lya$ forest power spectrum and the CMB lensing
convergence at $z=2.4$ (left) and 2.8 (right) assuming $\Delta z=0.2$, which
corresponds respectively to $\Delta r=166.63$ and $142.08\ihMpc$. $l$ is the
angular scale between the skewer and the lensing convergence, and $k_{s\para}$
is the scale of the one-dimensional $\lya$ forest power spectrum.}
\label{fig:true_sig}
\end{figure}

To evaluate the correlation between the one-dimensional forest power spectrum
and the CMB lensing convergence, we take the product of $C_l^{\bar\d\kappa}$ from
the left panel of \reffig{corr_bardkappa} and $dP^{\rm 1d}_{FF}(k_{s\para})/d\bar\d$
from the right panel of \reffig{pff1d}. \refFig{true_sig} shows the result at
$z=2.4$ (left) and 2.8 (right) as a function of $\ell$ (the angular scale between
the skewer and the lensing convergence) and $k_{s\para}$ (the scale of the one-dimensional
forest power spectrum), assuming $\Delta z=0.2$, which corresponds respectively
to $\Delta r=166.63$ and $142.08\ihMpc$. We find that at both redshifts the scale
dependence on $l$ and $k_{s\para}$ is similar, and the more apparent difference
is the amplitude of the signal. Specifically, at lower redshift the signal is
larger, which is likely due to the stronger gravitational evolution.

Lastly, if the correlation between the one-dimensional forest power spectrum
($\vr_{s\perp}=0$) and the lensing convergence is measured in configuration
space, then we can also Fourier transform the \emph{large-scale} $\vk_{l\perp}$:
\ba
 \langle P^{\rm 1d}_{FF}(k_{s\para},\vr_\perp)\kappa(\vr'_\perp)\rangle
 \:&=\int\frac{d^2 \vk_{l\perp}}{(2\pi)^2}
 \langle P^{\rm 1d}_{FF}(k_{s\para},\vk_{l\perp})\kappa(\vk_{l\perp})\rangle
 e^{i\vk_{l\perp}\cdot(\vr_\perp-\vr'_\perp)} \vs
 \:&=\frac{dP^{\rm 1d}_{FF}(k_{s\para})}{d\bar\d}\int\frac{d^2 \vk_{l\perp}}{(2\pi)^2}
 P^{\rm 2d}_{\bar\d\kappa}(\vk_{l\perp})e^{i\vk_{l\perp}\cdot(\vr_\perp-\vr'_\perp)} \,.
\ea
Furthermore, if the correlations are only measured at the same angular position
($\vr_\perp=\vr'_\perp$), as done in Ref.~\cite{Doux:2016xhg}, then the relevant
quantity in this case is the Fourier transform of $P^{\rm 2d}_{\bar{\delta}\kappa}$
evaluated at zero lag, i.e.
\be
 \sigma^2_{\bar\d\kappa}=\int\frac{d^2 \vk_{l\perp}}{(2\pi)^2}
 P^{\rm 2d}_{\bar\d\kappa}(\vk_{l\perp}) \,,
\ee
which the variance of the $\bar{\delta}\kappa$ field. Note that while this is the
region where majority of signal to noise is concentrated, there is in principle
more information available in the data.

We show the correlation at the same angular position in the right panel of
\reffig{corr_bardkappa} as a function of redshift for various $\Delta z$.
We confirm that the results are insensitive to the choice of $k_{\rm max}$
for the integration and the redshift evolution of $P_{\d\d}$. As for
$C_l^{\bar\d\kappa}$, $\sigma^2_{\bar\d\kappa}$ is larger at lower redshift
due to the gravitational evolution. Interestingly, while the comoving distances
between $\Delta z=0.1$ and 0.3 differ by a factor of three, the change in
$\sigma^2_{\bar\d\kappa}$ is only from 12 to 18\%. Should we compute
\be
 \sigma^2_{\bar\d\bar\d}=\int\frac{d^3k_l}{(2\pi)^3}P_{\d\d}(k_l)
 \sinc^2\(\frac{k_{l\para}\Delta r}{2}\) \,,
\ee
we have $\sigma^2_{\bar\d\bar\d}(\Delta z=0.1)/\sigma^2_{\bar\d\bar\d}(\Delta z=0.3)\approx2.9$
between $2.2\le z\le3.0$. This is because $\sigma^2_{\bar\d\bar\d}$ is the
variance of $\bar\d$ for a parallel scale $\Delta r$, and the larger the
$\Delta r$ the smaller the variance. On the other hand, $\sigma^2_{\bar\d\kappa}$
includes the integral of lensing kernel along the line-of-sight, and the
larger the $\Delta r$ the larger the lensing contribution. These two effects
roughly cancel each other, as a result $\sigma^2_{\bar\d\kappa}$ is much less
affected by the width in the line-of-sight direction.

%%%%%%%%%%%%%%%%%%%%%%%%%%%%%%%%%%%%%%%%%%%%%%%%%%%
\section{Continuum-misestimation bias due to Lyman-$\alpha$ forest measurement}
\label{sec:biased_signal}
In the previous section we compute the gravitational bispectrum of flux-flux-lensing.
In this section we shall discuss a non-standard systematic term coming from
the mis-estimation of the mean flux over the finite length of $\lya$ forest skewers,
which we shall refer to as the ``continuum-misestimation bias''.

Intuitively, this effect comes from the fact that over the length of
the forest, one cannot distinguish between a large-scale absorption
that would lower the measured flux in all pixels and the change in the
brightness of the quasars that would do exactly the same. Of course,
using information outside the quasar allows one to, in principle,
recover some of this information, but in practice this introduces
large amounts of noise in the continuum calibration. Therefore, most
analysis just \emph{fit} for the quasar brightness within the forest,
in effect nulling large-scale flux fluctuations modes, which leads to
two effects: i) distortion of measured large-scale power spectra or
correlation functions (see e.g. \cite{Slosar:2011mq,Blomqvist:2015pza})
and ii) modulation of the local small scale power spectrum normalization
by the large scale mode \cite{McDonald:2004eu}. The second effect is
the one that affects the calculation in this paper as we discuss next.

The observed $\lya$ flux can be modeled as
\be
 F(r_\para,\vr_\perp)=A(\vr_\perp)\bar{C}\bar{F}(r_\para)\[1+\delta_F(r_\para,\vr_\perp)\]+\epsilon \,,
\ee
where $A$ describes the brightness of the quasar so depends on the angular
position, $\bar{C}$ is the mean continuum shape obtained from stacking all
skewers in the survey, $\bar{F}$ is the true mean flux which is independent
of the angular position on the sky, $\delta_F$ is the true flux fluctuation
around $\bar{F}$, and $\epsilon$ is the noise. To measure $\delta_F$, one
performs the continuum fitting on a skewer by adjusting $A$ until the integral
of $\delta_F$ becomes zero across this skewer. In other words, for a skewer
at $\vr_\perp$ we estimate the quasar brightness as
\ba
 \hat{A}(\vr_\perp)\:&=\frac{1}{\Delta\tr}\int_{r_\para-\Delta\tr/2}^{r_\para+\Delta\tr/2}dr'_\para
 \frac{F(r'_\para,\vr_\perp)}{\bar{C}(r'_\para)\bar{F}(r'_\para)}
 =A(\vr_\perp)\[1+\frac{1}{\Delta\tr}\int_{r_\para-\Delta\tr/2}^{r_\para+\Delta\tr/2}
 dr'_\para\delta_F(r'_\para,\vr_\perp)\] \vs
 \:&=A(\vr_\perp)\[1+\Delta_F(r_\para,\vr_\perp,\Delta\tr)\] \,,
\ea
where $r_\para$ and $\Delta\tr$ are the center and the width of the skewer.
Note that the typical size of $\Delta\tr$ is 280 to $420\hMpc$ \cite{Doux:2016xhg},
so $\Delta\tr$ is generally larger than the size of the redshift bin $\Delta r$.
Plugging the estimated quasar luminosity $\hat A$ into the flux, we find the
estimated flux perturbation at $\vr_\perp$ to be (ignoring the $r_\para$ dependence
in $\Delta_F$ as it only describes the center of the forest)
\ba
 \hat{\d}_F(r'_\para,\vr_\perp)
 \:&=\frac{F(r'_\para,\vr_\perp)}{\hat{A}(\vr_\perp)\bar{C}(r'_\para)\bar{F}(r'_\para)}-1
 =\frac{1+\delta_F(r'_\para,\vr_\perp)}{1+\Delta_F(\vr_\perp,\Delta\tr)}-1 \vs
 \:&\approx\delta_F(r'_\para,\vr_\perp)\[1-\Delta_F(\vr_\perp,\Delta\tr)\]
 -\Delta_F(\vr_\perp,\Delta\tr) \,.
\label{eq:hat_dF}
\ea
Note that the last term in \refeq{hat_dF} is a constant at this angular position
of $\vr_\perp$, hence it contributes only to $\vk_s=0$ mode of $\hat{\d}_F$ in
Fourier space. We shall ignore it as we are interested in the forest power spectrum
with $k_s>0$, and to the leading order the estimated forest power spectrum of this
skewer becomes
\be
 \hat{P}_{FF}(\vk_s,\vr_\perp)=\langle\hat{\d}_F(\vk_s,\vr_\perp)\hat{\d}_F(\vk'_s,\vr_\perp)\rangle'
 =P_{FF}(\vk_s)\[1-2\Delta_F(\vr_\perp,\Delta\tr)\] \,,
\ee
where the ensemble average is taken for the small-scale mode $\vk_s$ while keeping
the large-scale $\Delta_F$ fixed. Physically, the correction term appears because
we use the {\it local} instead of the {\it global} mean flux (which is independent
of $\vr_\perp$) to compute the fluctuation. The same effect is also discussed in
Ref.~\cite{Chiang:2015eza} for the measurement of the position-dependent correlation
function. Moreover, the larger the $\Delta\tr$ the smaller the correction.

\begin{figure}[t]
\centering
\includegraphics[width=0.495\textwidth]{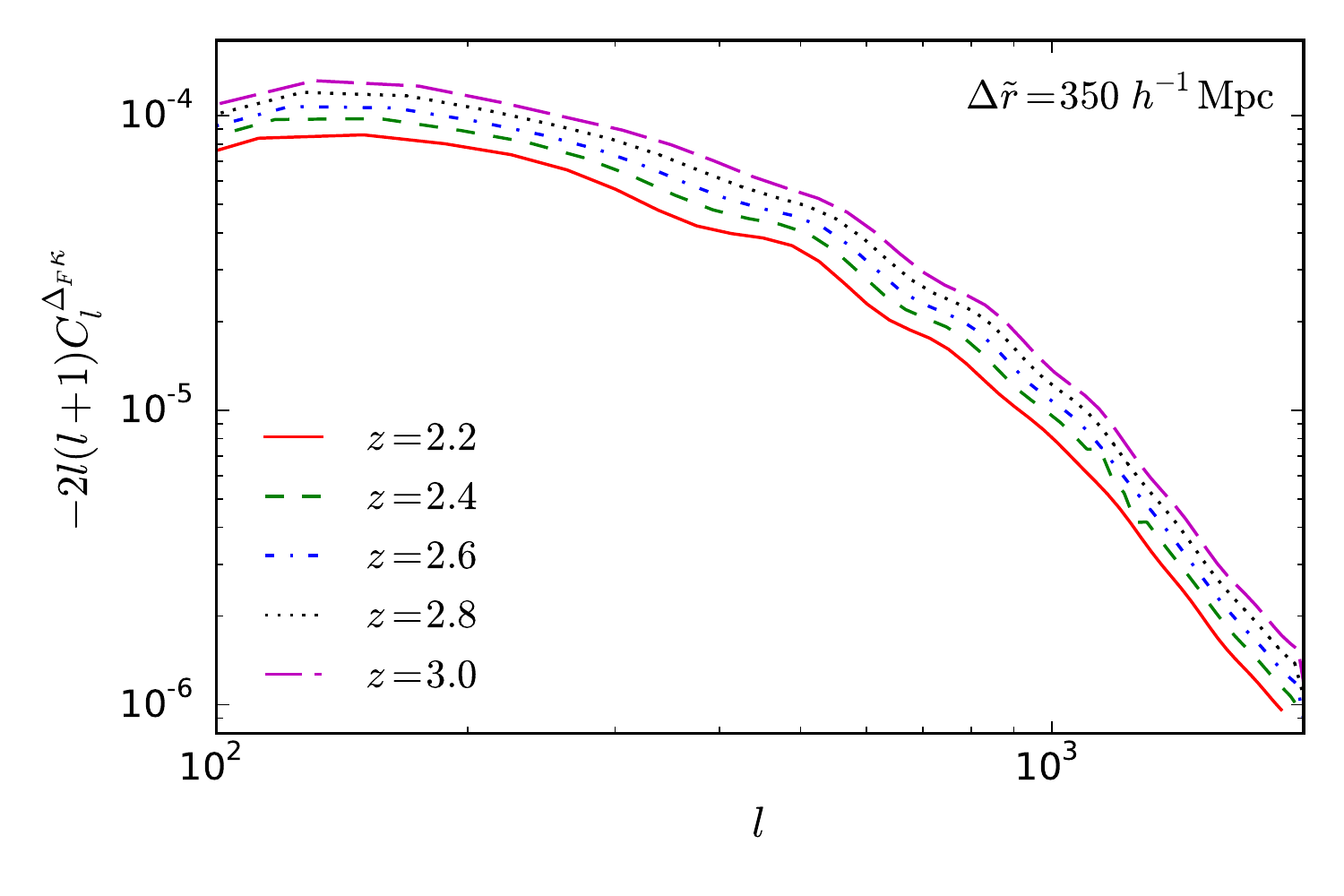}
\includegraphics[width=0.495\textwidth]{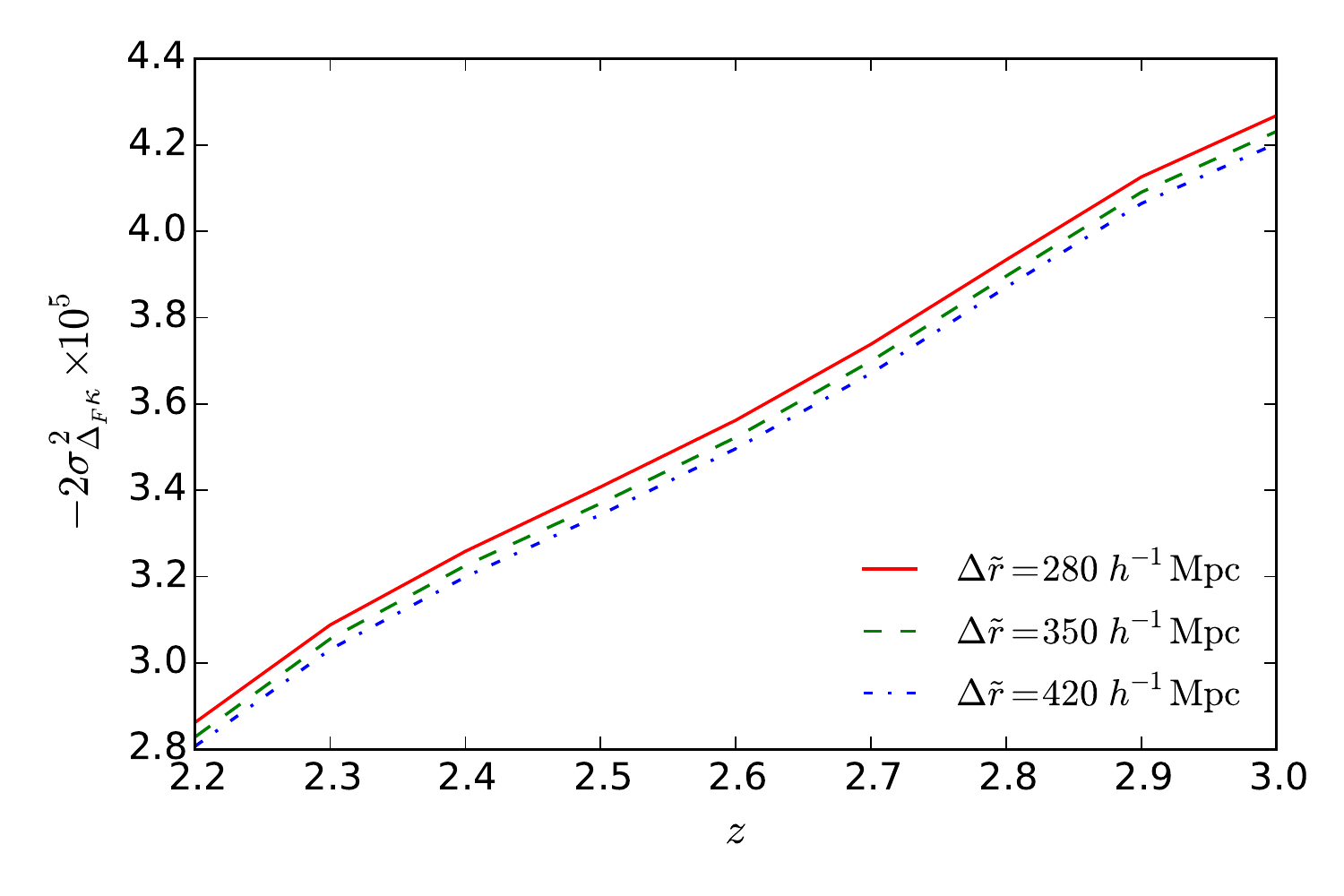}
\caption{(Left) The angular power spectrum of the continuum-misestimation bias $C_l^{\Delta_F\kappa}$
with the length of the forest $\Delta\tr=350\hMpc$ as a function of the angular
scale $l$ at various redshifts denoted by different colors and line styles. (Right)
The correlation between $\Delta_F$ and $\kappa$ at the same angular position,
$\sigma^2_{\Delta_F\kappa}$, as a function of redshift for various lengths of
the forest $\Delta\tr$ denoted by different colors and line styles.}
\label{fig:corr_DFkappa}
\end{figure}

Though for each skewer the estimated forest power spectrum is biased,
the total estimated forest power spectrum is still unbiased at the first
order since $\left<\Delta_F\right>=0$. It is well known that there are
corrections at the second order \cite{McDonald:2004eu}. However, if we
measure the correlation between the forest power spectrum of each skewer
and the CMB lensing convergence, the signal would be biased due to the
correlation between $\Delta_F$ and $\kappa$ at the first order. Specifically,
if we consider the angular power spectrum between the forest power spectrum
and the lensing convergence, then the continuum-misestimation bias is
\be
 \langle\hat{P}_{FF}(\vk_s,\vk_{l\perp})\kappa(\vk'_{l\perp})\rangle'
 =-2P_{FF}(\vk_s)\langle\Delta_F(\vk_{l\perp},\Delta\tr)\kappa(\vk'_{l\perp})\rangle'
 =-2P_{FF}(\vk_s)P^{\rm 2d}_{\Delta_F\kappa}(\vk_{l\perp}) \,,
\ee
where
\be
 P^{\rm 2d}_{\Delta_F\kappa}(\vk_{l\perp})
 =\int_{r_\para-\Delta\tr/2}^{r_\para+\Delta\tr/2}dr'_\para W_\kappa(r'_\para)
 \int\frac{dq_\para}{2\pi}P_{\d F}(q_\para,\vk_{l\perp})\Lambda_\kappa(-\vk_{l\perp})
 \sinc\(\frac{q_\para\Delta\tr}{2}\)e^{iq_\para(r'_\para-r_\para)} \,,
\label{eq:p2d_DFkappa}
\ee
with $P_{\d F}(\vk)=\[P_{\d\d}(\vk)P_{FF}(\vk)\]^{1/2}$ in linear theory.
Note that \refeq{p2d_DFkappa}
is basically the same as \refeq{p2d_bardkappa}, excpet for different power
spectra as well as the lengths appearing in the integration boundary and
the sinc function. The left panel of \reffig{corr_DFkappa} shows the angular
power spectrum of the continuum-misestimation bias $C_l^{\Delta_F\kappa}$
under the flat-sky approximation with $\Delta\tr=350\hMpc$ as a function
of the angular scale $l$ at various redshifts. Unlike $C_l^{\bar\d\kappa}$
shown in \reffig{corr_bardkappa}, we find that the continuum-misestimation
bias is larger at higher redshift, which is due to the fact that $C_l^{\Delta_F\kappa}$
is proportional to $b_F$ and $|b_F|$ is larger at higher redshift.

\begin{figure}[t]
\centering
\includegraphics[width=0.495\textwidth]{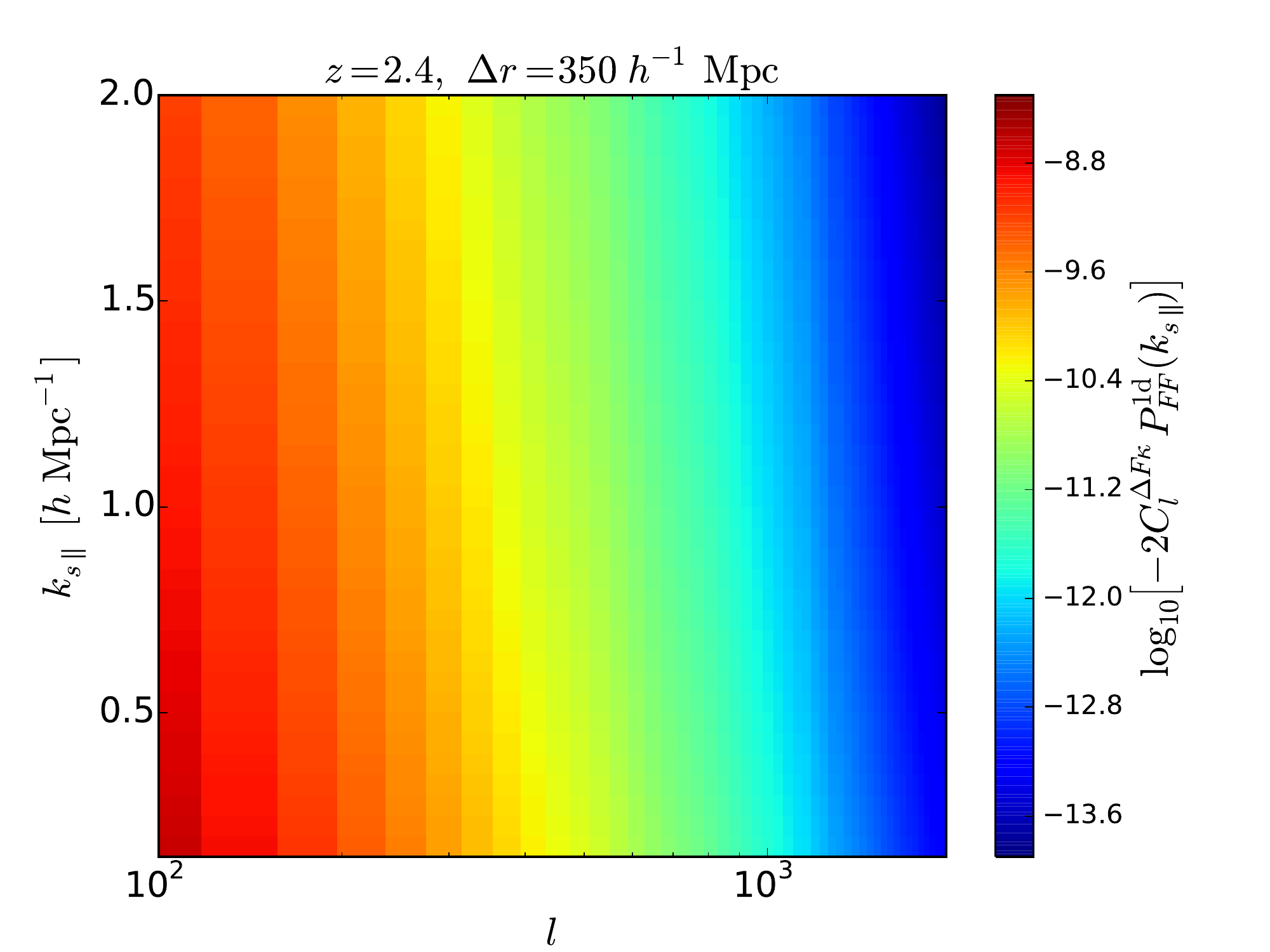}
\includegraphics[width=0.495\textwidth]{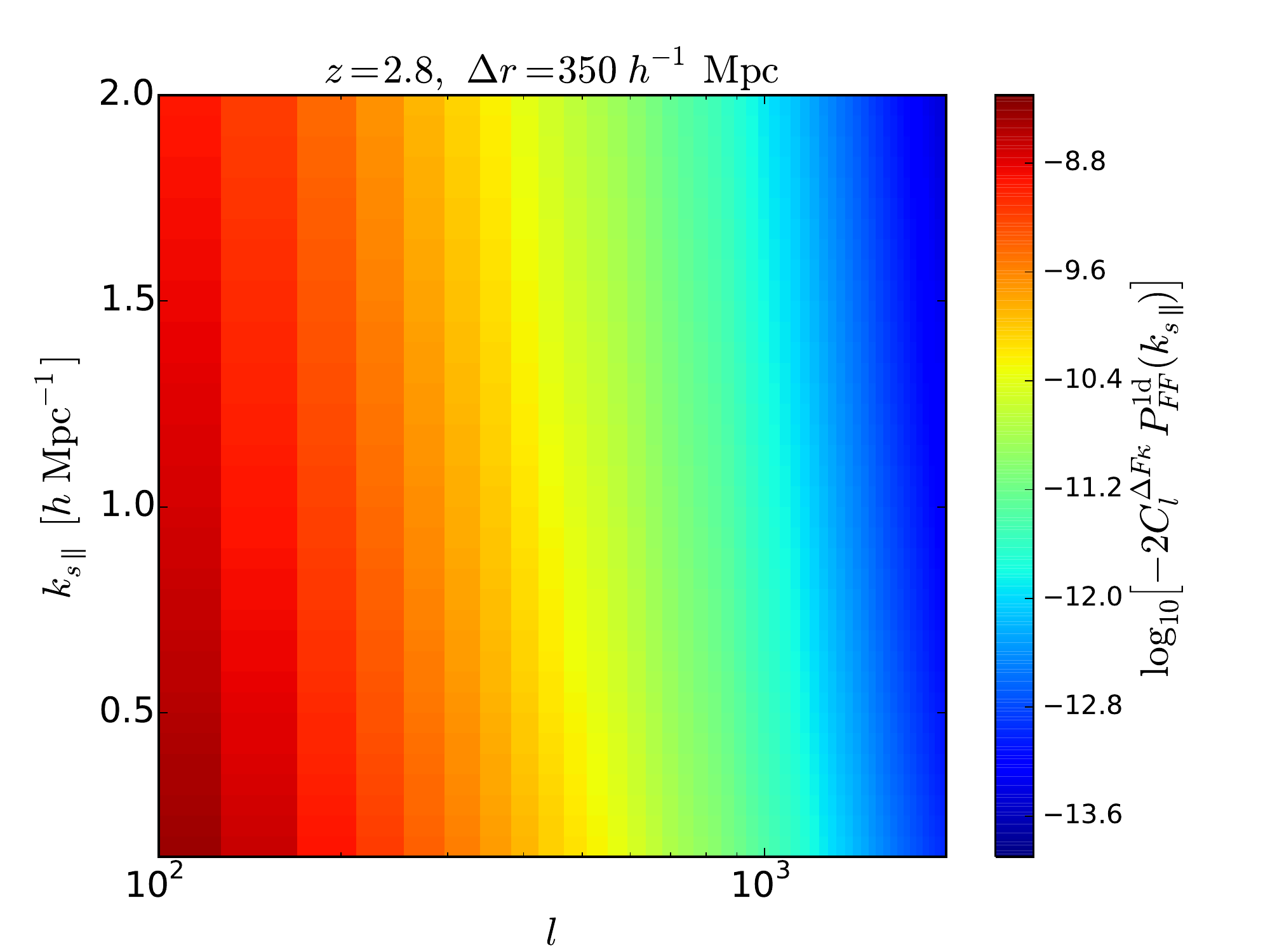}
\caption{Continuum-misestimation bias of the correlation between the one-dimensional
$\lya$ forest power spectrum and the lensing convergence due to the continuum
fitting at $z=2.4$ (left) and 2.8 (right), assuming the length of the
forest to be $350\hMpc$. The bias signal is plotted as a function of $l$,
which is the angular scale between the skewer and the lensing convergence,
and $k_{s\para}$, which is the scale of the one-dimensional $\lya$ forest power
spectrum.}
\label{fig:bias_sig}
\end{figure}

For the correlation between the one-dimensional forest power spectrum and the
lensing convergence, the continuum-misestimation bias becomes $-2P^{\rm 1d}_{FF}(\vk_s)P^{\rm 2d}_{\Delta_F\kappa}(\vk_{l\perp})$.
To evaluate the signal, we take the one-dimensional forest power spectrum
from the left panel of \reffig{pff1d} and $C_l^{\Delta_F\kappa}$ from the right
panel of \reffig{corr_DFkappa}, and the result is shown in \reffig{bias_sig} at
$z=2.4$ (left) and 2.8 (right), assuming the length of the forest to be $350\hMpc$.
We find that the scale dependences of the true signal and the continuum-misestimation bias are similar.
This is because the scale dependences are predominantly determined by $C_l^{\bar\d\kappa}$
and $C_l^{\Delta_F\kappa}$, and the primary difference between the two angular
power spectra is the flux bias as on large scale $\dnl\to1$. More interestingly,
the relative contribution of the continuum-misestimation bias becomes larger at higher redshift,
which is the result of increasing $|b_F|$ at higher redshift.

Finally, for the correlation measured in configuration space at the same angular
position, the continuum-misestimation bias is $-2P^{\rm 1d}_{FF}(k_{s\para})\sigma^2_{\Delta_F\kappa}$,
where $\sigma^2_{\Delta_F\kappa}$ is the Fourier transform of $P^{\rm 2d}_{\Delta_F\kappa}(\vk_{l\perp})$
at zero lag, i.e. 
\be
 \sigma^2_{\Delta_F\kappa}=\int\frac{d^2k_{l\perp}}{(2\pi)^2}P^{\rm 2d}_{\Delta_F\kappa}(\vk_{l\perp}) \,.
\ee
The right panel of \reffig{corr_DFkappa} shows $\sigma^2_{\Delta_F\kappa}$ as
a function of redshift for various $\Delta\tr$. We find that as the left panel
the continuum-misestimation bias is larger at high redshift due to the larger bias. Also,
$\sigma^2_{\Delta_F\kappa}$ is almost unchanged between $\Delta\tr=280$ to
$420\hMpc$. This is not surprising because the integration over the lensing
kernel cancels the effect from the variance of the density perturbation
$\sigma^2_{\bar\d\bar\d}$, as we have discussed in \refsec{true_signal}.
It is, however, interesting in the sense that in a $\lya$ survey the lengths
of the forest may vary a lot, but to model the continuum-misestimation bias
of the lensing-flux-flux bispectrum we only need one typical length for the
forest, hence the modeling is simpler.

%%%%%%%%%%%%%%%%%%%%%%%%%%%%%%%%%%%%%%%%%%%%%%%%%%%
\section{Comparison with observation}
\label{sec:observation}
How does our prediction compare to the measurement? In this section we shall
first discuss the contamination from damped $\lya$ systems that cannot be
removed perfectly in observation, and then compare our prediction to the
measured correlation between the one-dimensional forest power spectrum and
CMB lensing convergence done in Ref.~\cite{Doux:2016xhg}.

\subsection{Contamination from Damped Lyman-$\alpha$ systems}
In the $\lya$ forest measurement, the largely broadened damping wings produced by
high neutral hydrogen column density systems with $N_{HI}>2\times10^{20}~{\rm cm}^{-2}$,
which is known as the damped $\lya$ absorbers (DLAs), are discarded to avoid contamination
on the forest power spectrum. However, systems such as Lyman-limit systems and sub-DLAs
with less column density than DLAs but still higher than that of the $\lya$ forest
have less prominent features hence are difficult to remove. The presence of these
systems may affect the forest power spectrum and so its correlation with the
lensing convergence, and we shall quantify the effect.

As pointed out in Ref.~\cite{McDonald:2004xp}, the dominant effect of the high
column density systems is to add power to the forest power spectrum due to their
random distribution. In the limit that the number of the systems are small (typically
in one skewer one in 100-1000 pixels is in the high column density system), the
power is proportional to the number density of such systems. As a result, in the
presence of a large-scale fluctuation $\bar\d$, the leading order correction to
the one-dimensional forest power spectrum due to the high column density systems
is given by
\be
 P^{\rm 1d}_{\rm DLA}(k_{s\para},\vr_\perp)=
 \[1+b_{\rm DLA}\bar\d(\vr_\perp)\]P^{\rm 1d}_{\rm DLA}(k_{s\para}) \,,
\ee
where $P^{\rm 1d}_{\rm DLA}$ and $b_{\rm DLA}$ are the one-dimensional power
spectrum and the bias of the high column density systems. Correlating this with
the lensing convergence, the general three-point function containing the angular
scale between the small-scale one-dimensional DLA power spectrum and the lensing
convergence is
\be
 \langle P^{\rm 1d}_{\rm DLA}(k_{s\para},\vk_{l\perp})\kappa(\vk'_{l\perp})\rangle'
 =b_{\rm DLA}P^{\rm 1d}_{\rm DLA}(k_{s\para})P^{\rm 2d}_{\bar\d\kappa}(\vk_{l\perp}) \,.
\ee
If the correlation is computed as the same angular position, then the signal
becomes
\be
 \langle P^{\rm 1d}_{\rm DLA}(k_{s\para},\vr_\perp)\kappa(\vr_\perp)\rangle
 =b_{\rm DLA}P^{\rm 1d}_{\rm DLA}(k_{s\para})\sigma^2_{\bar\d\kappa} \,.
\ee

To evaluate this signal, we need both the bias and the one-dimensional power
spectrum of DLAs. Ref.~\cite{FontRibera:2012fm} has measured $b_{\rm DLA}$
by cross-correlating DLAs with $\lya$ forest, and we should take their central
value $b_{\rm DLA}=2.17$, assuming that the redshift-space distortion parameter
of DLAs is subdominant. For the DLA power spectrum, we use the fitting function
provided in Ref.~\cite{Rogers:2017bmq} describing the ratio between $P^{\rm 1d}_{\rm DLA}$
and $P^{\rm 1d}_{FF}$, and we utilize our fiducial forest power spectrum. Note
that we only consider the contributions from Lyman-limit systems and sub-DLAs
(i.e. the red line in figure~5 of Ref.~\cite{Rogers:2017bmq}), as the prominent
DLAs are discarded in Ref.~\cite{Doux:2016xhg}.

\subsection{Measurement of Doux et al.}
In Ref.~\cite{Doux:2016xhg}, the first detection of the correlation between the
one-dimensional forest power spectrum from SDSS-III Baryon Oscillation Spectroscopic
Survey Data Release 12 \cite{Dawson:2012va,Paris:2016xdm} and the CMB lensing
convergence from Planck \cite{Ade:2015zua} at the same line-of-sight is reported
with a significance of 5 $\sigma$. The authors interpreted this signal to be the
response of the forest power spectrum to a large-scale overdensity, and subtracting
the contribution from the leading-order squeezed-limit matter bispectrum they
measured the ``effective nonlinear bias'', which contains the nonlinearity such
as redshift-space distortion and nonlinear clustering of gas. We shall compare the
measurement of $\langle P^{\rm 1d}_{FF}(k_{s\para},\vr_\perp)\kappa(\vr_\perp)\rangle$
with our prediction.

Let us first compute all components of the signal. The true signal is produced by
the nonlinear gravitational evolution and given by
$\[dP^{\rm 1d}_{FF}(k_{s\para})/d\bar\d\]\sigma^2_{\bar\d\kappa}$, where the
$dP^{\rm 1d}_{FF}(k_{s\para})/d\bar\d$ and $\sigma^2_{\bar\d\kappa}$ are taken from
the right panel of \reffig{pff1d} and the right panel of \reffig{corr_bardkappa},
respectively. The continuum-misestimation bias is due to the $\lya$ forest continuum
fitting and given by $-2P^{\rm 1d}_{FF}(k_{s\para})\sigma^2_{\Delta_F\kappa}$, where
$P^{\rm 1d}_{FF}(k_{s\para})$ and $\sigma^2_{\Delta_F\kappa}$ are taken from the left
panel of \reffig{pff1d} and the right panel of \reffig{corr_DFkappa}. Finally, the DLA
contamination is given by $b_{\rm DLA}P^{\rm 1d}_{\rm DLA}(k_{s\para})\sigma^2_{\bar\d\kappa}$,
where $b_{\rm DLA}=2.17$ and $P^{\rm 1d}_{\rm DLA}(k_{s\para})$ is taken from the fitting
function in Ref.~\cite{Rogers:2017bmq}. Thus, we predict the total signal of the
correlation between the one-dimensional forest power spectrum and the lensing convergence
from the same line-of-sight to be
\be
 \frac{dP^{\rm 1d}_{FF}(k_{s\para})}{d\bar\d}\sigma^2_{\bar\d\kappa}
 -2P^{\rm 1d}_{FF}(k_{s\para})\sigma^2_{\Delta_F\kappa}
 +b_{\rm DLA}P^{\rm 1d}_{\rm DLA}(k_{s\para})\sigma^2_{\bar\d\kappa} \,.
\ee

\begin{figure}
\centering
\hspace*{-0.5cm}
\begin{tabular}{cc}
\includegraphics[width=0.5\textwidth]{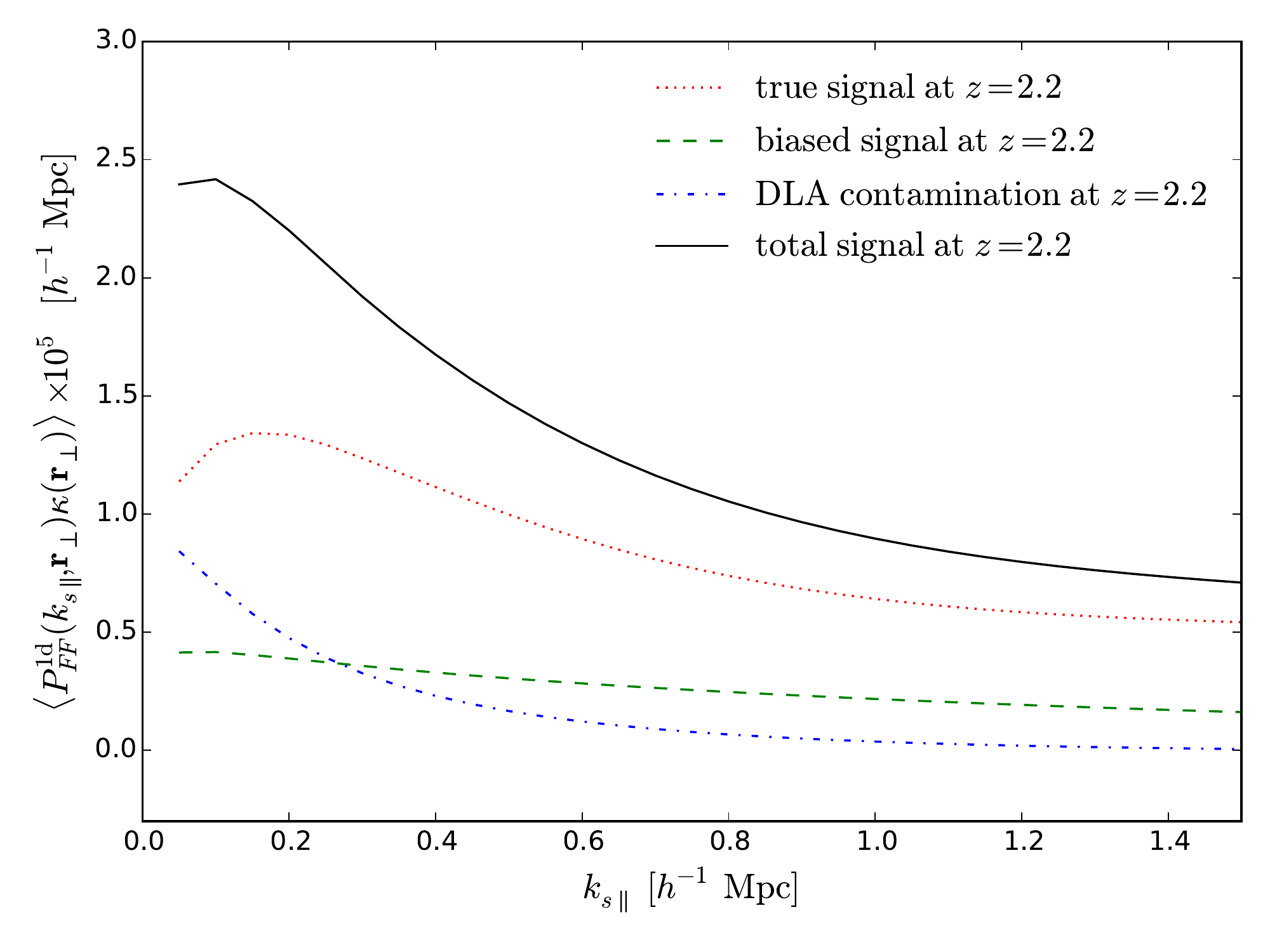} &
\includegraphics[width=0.5\textwidth]{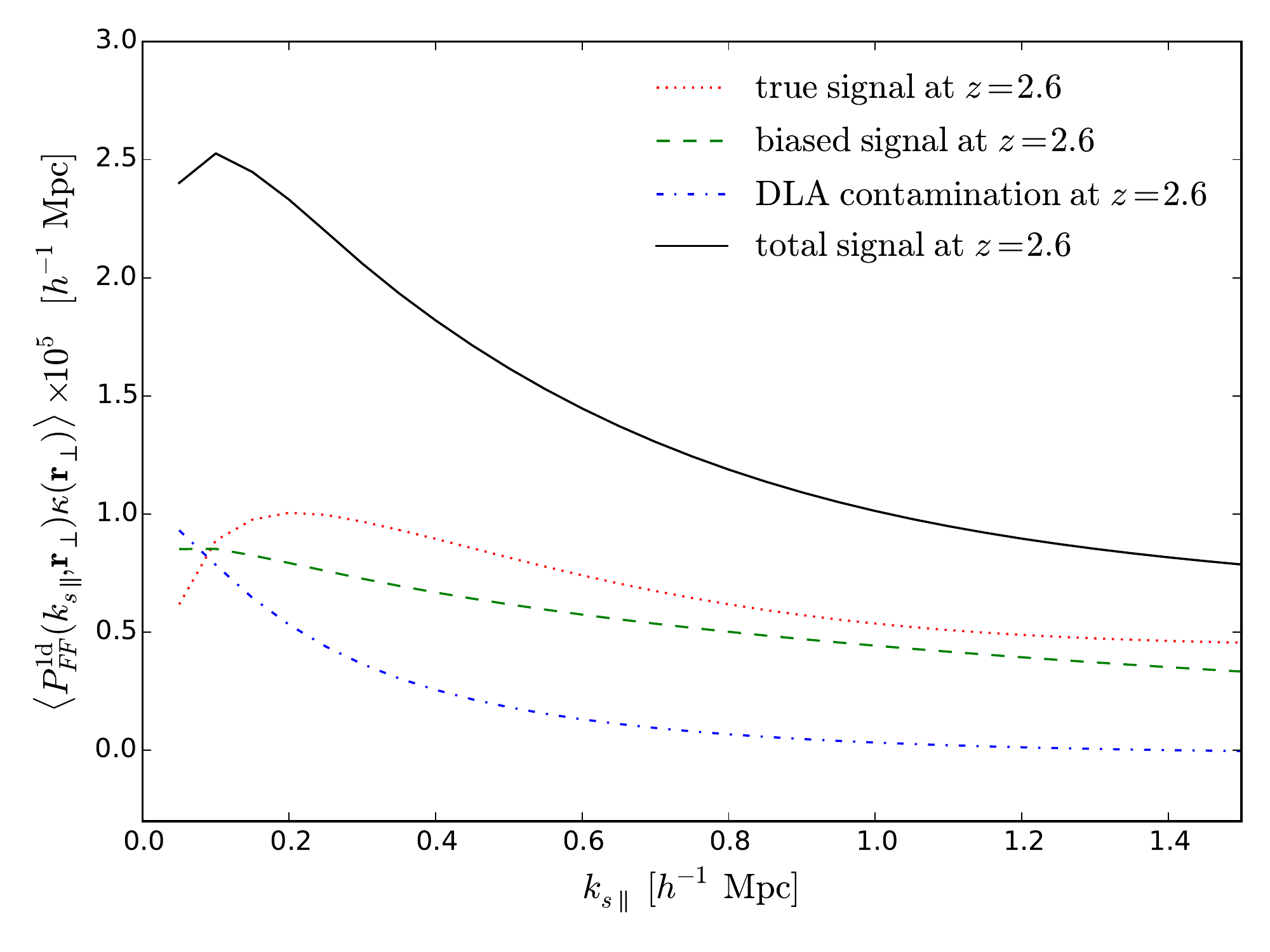} \\
\includegraphics[width=0.5\textwidth]{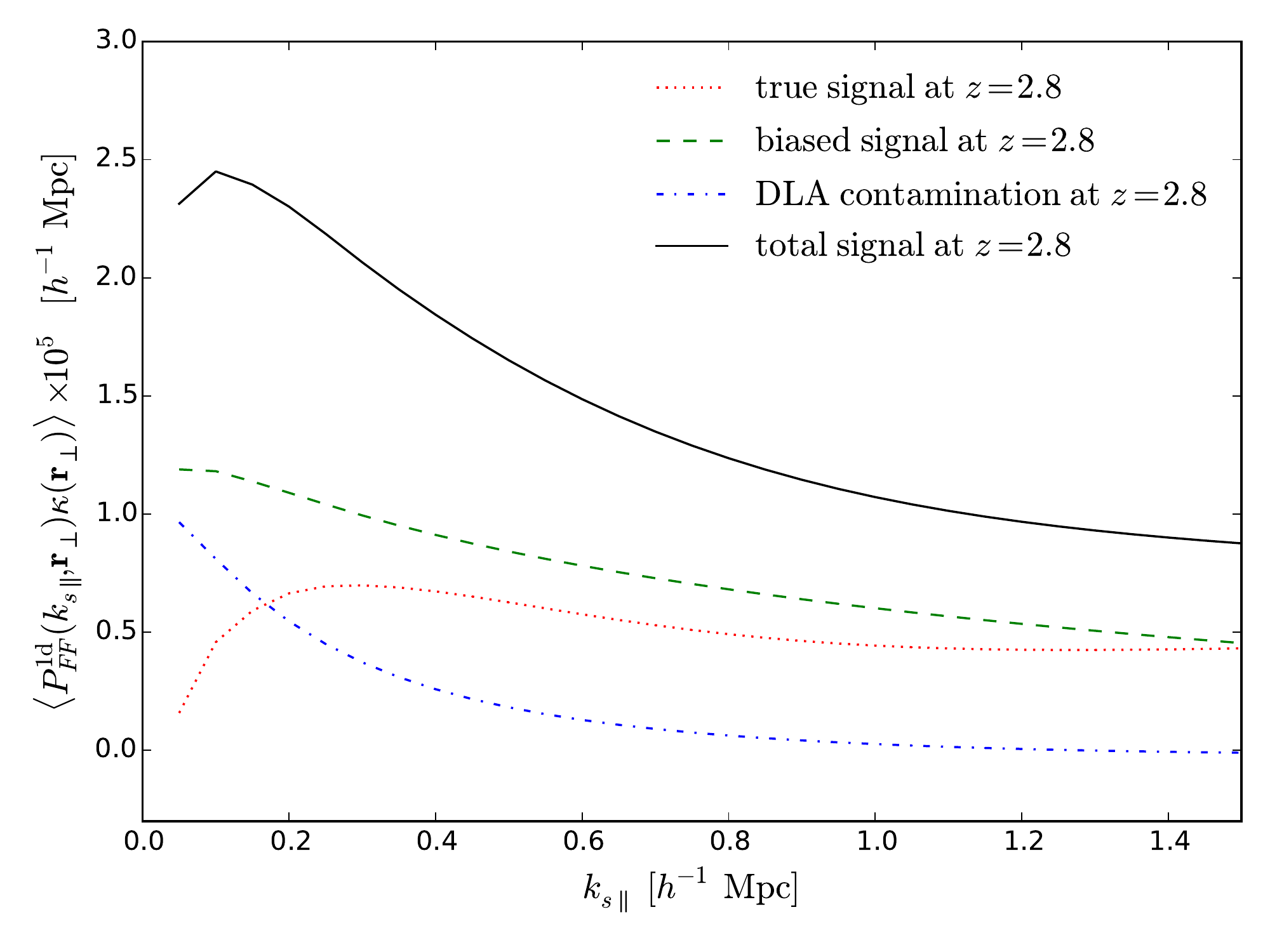} &
\includegraphics[width=0.5\textwidth]{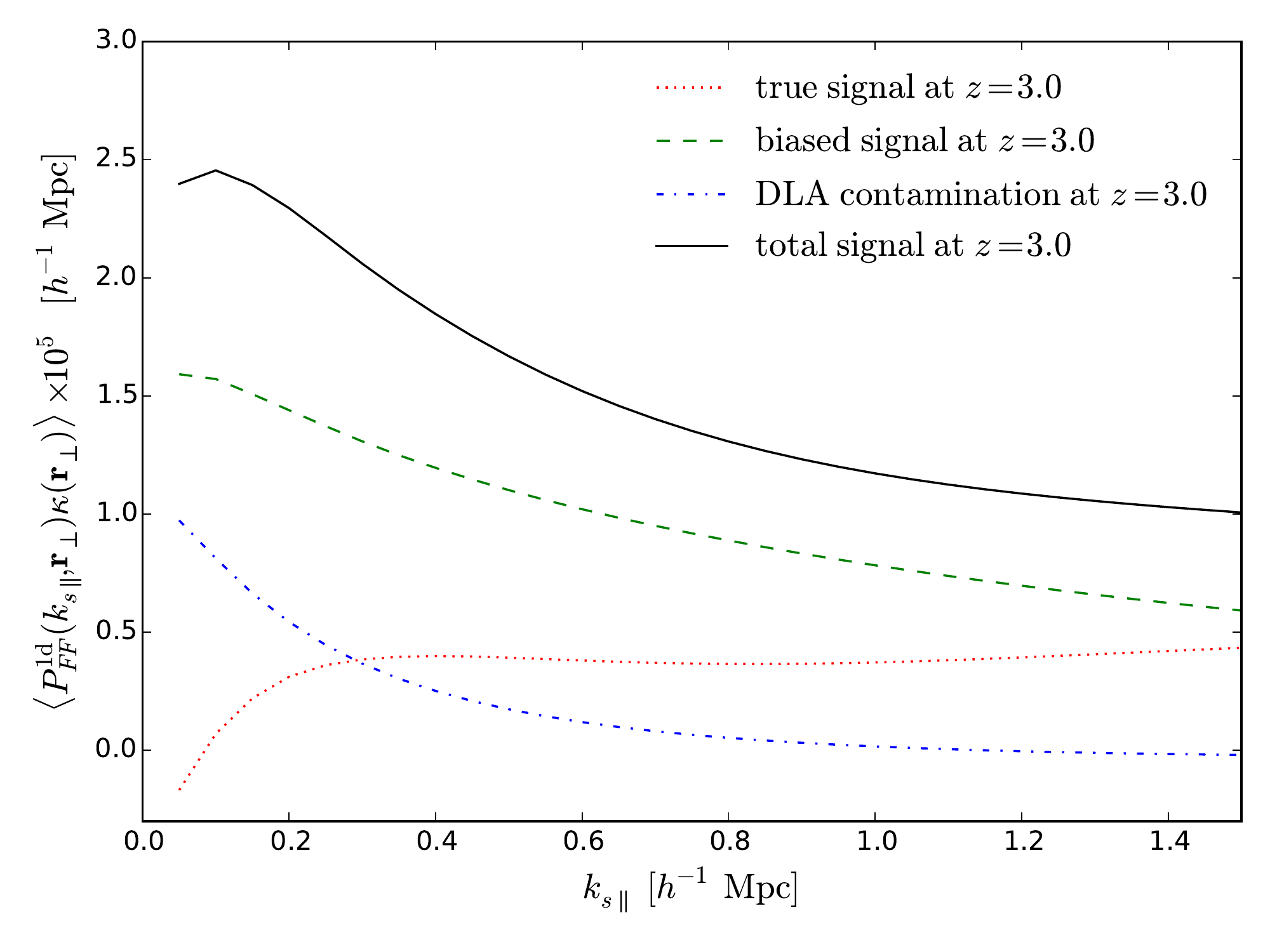}\\
\end{tabular}
\caption{Different components of the signal of the cross-correlation between lensing
convergence and the $\lya$ forest power spectrum at the same angular position, i.e.
$\langle P^{\rm 1d}_{\rm DLA}(k_{s\para},\vr_\perp)\kappa(\vr_\perp)\rangle$. The red
dotted line shows the true signal that is produced by the nonlinear gravitational evolution,
the green dashed line shows the continuum-misestimation bias due to the $\lya$ continuum fitting,
the blue dot-dashed line shows the DLA contamination, and the black solid line shows
the total signal. The redshift of the signal increases from the top left to bottom
right panels.}
\label{fig:sig_tot}
\end{figure}

\refFig{sig_tot} shows the different components of the cross-correlation between
lensing convergence and the forest power spectrum at the same angular position
for various redshifts. We first find that the DLA contamination is only important
on large scales, and for $k_{s\para}\gtrsim0.6~\ihMpc$ the contribution becomes
negligible. This is due to the shape of the DLA power spectrum, which can be seen
in figure~5 of Ref.~\cite{Rogers:2017bmq}. Next, we find that true signal dominates
over the continuum-misestimation bias at lower redshift, but the trend reverses at
high redshift. This is due to the increase of the $\lya$ flux bias, which enhances
both $P^{\rm 1d}_{FF}(k_{s\para})$ and $\sigma^2_{\Delta_F\kappa}$, as well as the
decrease of the forest power spectrum response. The fact that the continuum-misestimation
bias is larger than the true signal at $z\ge2.8$ means that a large portion of the
measured signal is coming from the mis-estimation of mean flux over the finite length
of $\lya$ forest skewers, which needs to be removed for unbiasedly extracting the
three-point function due to the gravitational evolution. Interestingly, the total signal
shows a mild redshift evolution between $z=2.2$ and 3.0. However, due to the lack
of snapshots of separate universe simulations at $z\ge3$ we cannot examine this
correlation at higher redshift where some of the forests are observed.

\begin{figure}
\centering
\includegraphics[width=0.8\textwidth]{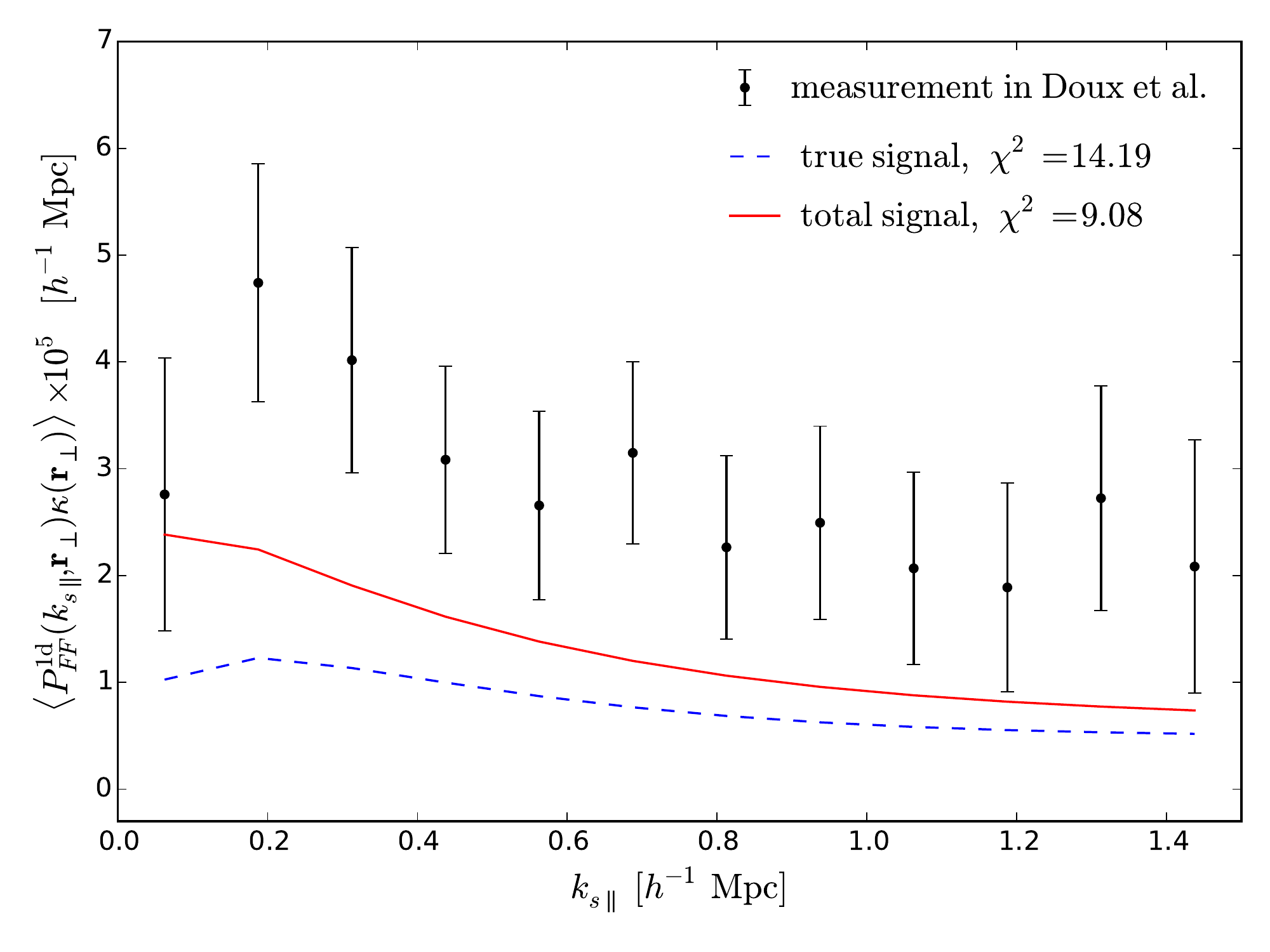}
\caption{Comparison of the cross-correlation between lensing convergence and the
$\lya$ forest power spectrum at the same angular position
$\langle P^{\rm 1d}_{\rm DLA}(k_{s\para},\vr_\perp)\kappa(\vr_\perp)\rangle$
between the measurement in Doux et al. (black data points with error bars) to
the true (blue dashed) and total signals (red solid). The values of $\chi^2$
computed using the full covariance matrix are shown in the legend with 12
degrees of freedom.}
\label{fig:meas}
\end{figure}

Since Ref.~\cite{Doux:2016xhg} measures the correlation from all
redshifts between $2.1\le z\le3.6$, we shall combine the total signals
at different redshifts in \reffig{sig_tot} into one line in order to
compare with the measurement. We use the weight given in eq.~(18) of
Ref.~\cite{Doux:2016xhg}, assuming that the number of pixels of the
$i^{\rm th}$ forest is a constant and noise of the forest is zero.  We
also include the observed $\lya$ forest distribution obtained from the
sample in Ref.~\cite{Doux:2016xhg}, hence the total weight is
\be
 w(k,z)=\frac{N_{\lya}(z)}{\[P^{\rm 1d}_{FF}(k,z)\]^2} \,,
\ee
where $N_{\lya}(z)$ is the number of $\lya$ forest at $z$.
The joint true and total signals from $2.2\le z\le3$ are shown
respectively as the blue dashed and red solid lines in \reffig{meas},
whereas the measurement in Ref.~\cite{Doux:2016xhg} is shown as the
black data points with error bar.  We find that the total signal is
closer to the measurement than the true signal alone, but still
smaller than the measurement on all scales. To better quantify the
difference, we compute $\chi^2$ using the full covariance matrix
estimated in Ref.~\cite{Doux:2016xhg}, and the values are shown in the
legend. As there are 12 data points and without any fitting
parameters, the degrees of freedom is 12. While the measurement is
consistently larger than the total signal, we still obtain a
reasonable reduced $\chi^2$. This is due to the high correlation
between measurement points, which is shown in figure~5 of
Ref.~\cite{Doux:2016xhg} and can be noticed by the lack of scatter
between neighboring points. We find that the total signal gives a
$\chi^2=9.1$ and that neglecting the biased signal and DLA
contamination worsens the fit to $\chi^2=14.19$. Therefore, even in
the current data, it is important to include new effects discussed in
this paper. We also note that Doux et al report best-fit $\chi^2$ of
5.4 (which, while low, is not anomalously low with $p=0.9$ with 11
degrees of freedom). Their model would roughly correspond to an
artificial increase in bias for our true signal, but the improvement
in fit is not significant, so we cannot exclude our model as a
complete model for the data.  One would be able to examine this more
critically with future surveys that have more and better $\lya$ forest
measurements.

%%%%%%%%%%%%%%%%%%%%%%%%%%%%%%%%%%%%%%%%%%%%%%%%%%%
\section{Conclusion}
\label{sec:conclusion}
In this paper, we investigate the three-point function form by the cross-correlation
between forest power spectrum and the CMB lensing convergence. We find that
not only the gravitational evolution produces the flux-flux-lensing bispectrum,
but the mis-estimation of the mean flux over the finite length of $\lya$ forest skewers
would also generate non-zero correlation. In particular, this systematic effect
dominates the underlying signal at $z\gtrsim2.8$, which has to be taken into
account for unbiasedly probing the gravitational effect at high redshift using
this specific bispectrum.

We demonstrate that integrating the flux-flux-lensing bispectrum with full
angular and scale dependences over the angular information between the forest
power spectrum and the CMB lensing is equivalently to the cross-correlation
between the forest power spectrum and the CMB lensing at the same angular
positions, which is measured in Ref.~\cite{Doux:2016xhg}. We show that our
predictions are consistent with the signal measured by Ref.~\cite{Doux:2016xhg},
and the inclusion of the systematic effects from $\lya$ forest continuum
fitting and the DLA contamination improves the agreement. However, for
both with and without the systematic effects, we obtain acceptable $\chi^2$
values. Furthermore, the reported best-fit $\chi^2$ with one fitting parameter
in Ref.~\cite{Doux:2016xhg} is low but not anomalous small.

If our prediction is correct, then one interesting implication is that while
Ref.~\cite{Doux:2016xhg} shows that the effective nonlinear bias, which encodes
the excess of the forest power spectrum response with respect to that of the
linear field, is $b_2^{\rm eff}=1.16\pm0.53$, our simulations suggest it to be
negative as the forest power spectrum response $d\ln P^{\rm 1d}_{FF}(k_{s\para})/d\delta$
is always smaller than the linear power spectrum response $d\ln P_l(k)/d\delta=68/21-(1/3)[d\ln k^3P_l(k)/d\ln k]$
for all redshifts and lines-of-sight (see figure~2 of Ref.~\cite{Chiang:2017vsq}).
This is likely due to the missing of the systematic term, hence to interpret
$b_2^{\rm eff}$ measured in Ref.~\cite{Doux:2016xhg} as the responses of flux
bias, redshift-space distortion, and the nonlinear small-scale forest power
spectrum one has to account for the continuum-misestimation bias. However,
given the large uncertainty from current measurement, it is challenging to
draw any concrete conclusions. With future surveys such as DESI \cite{Aghamousa:2016zmz}
that have less uncertainty, one can more critically examine this correlation
as well as how the forest power spectrum responds to the overdensity gravitationally.

\acknowledgments We would like to thank Cyrille Doux for providing the
data points and the covariance matrix of the measurement in Ref.~\cite{Doux:2016xhg}.
We would also like to thank Emmanuel Schaan and the referees for useful
comments on the draft. AS acknowledges insightful conversations
with Andreu Font-Ribera and hospitality of the University College
London where parts of this work were performed. Results in this paper
were obtained using the high-performance computing system at the
Institute for Advanced Computational Science at Stony Brook
University. CC is supported by grant NSF PHY-1620628.

%%%%%%%%%%%%%%%%%%%%%%%%%%%%%%%%%%%%%%%%%%%%%%%%%%%%%%%%%%%%%%%%%%%%%%%%%%%%

\bibliography{draft}
\end{document}